\documentclass[lettersize,journal]{IEEEtran}
\usepackage{amsmath,amsfonts}
\usepackage{algorithmic}
\usepackage{algorithm}
\usepackage{array}
\usepackage[caption=false,font=normalsize,labelfont=sf,textfont=sf]{subfig}
\usepackage{textcomp}
\usepackage{stfloats}
\usepackage{url}
\usepackage{verbatim}
\usepackage{graphicx}
\usepackage{cite}
\usepackage{amssymb}
\usepackage{multirow}
\usepackage{booktabs}
\usepackage{xcolor}

\begin{document}

\title{Frequency-Spatial Interaction Driven Network for Low-Light Image Enhancement}

\author{Yunhong Tao, Wenbing Tao$^*$, Xiang Xiang$^*$
\thanks{* Yunhong Tao, Wenbing Tao, and Xiang Xiang are all with School of Artificial Intelligence and Automation, while Xiang Xiang is also with School of Computer Science and Technology, both at Huazhong University of Science and Technology, China. Corresponding co-author: Wenbing Tao, Xiang Xiang.}
}

\markboth{Preprinted Technical Report}%
{Shell \MakeLowercase{\textit{et al.}}: A Sample Article Using IEEEtran.cls for IEEE Journals}


\maketitle

\begin{abstract}
Low-light image enhancement (LLIE) aims at improving the perception or interpretability of an image captured in an environment with poor illumination. With the advent of deep learning, the LLIE technique has achieved significant breakthroughs. However, existing LLIE methods either ignore the important role of frequency domain information or fail to effectively promote the propagation and flow of information, limiting the LLIE performance. In this paper, we develop a novel frequency-spatial interaction-driven network (FSIDNet) for LLIE based on two-stage architecture. To be specific, the first stage is designed to restore the amplitude of low-light images to improve the lightness, and the second stage devotes to restore phase information to refine fine-grained structures. Considering that Frequency domain and spatial domain information are complementary and both favorable for LLIE, we further develop two frequency-spatial interaction blocks which mutually amalgamate the complementary spatial and frequency information to enhance the capability of the model. In addition, we construct the Information Exchange Module (IEM) to associate two stages by adequately incorporating cross-stage and cross-scale features to effectively promote the propagation and flow of information in the two-stage network structure. Finally, we conduct experiments on several widely used benchmark datasets (i.e., LOL-Real, LSRW-Huawei, etc.), which demonstrate that our method achieves the excellent performance in terms of visual results and quantitative metrics while preserving good model efficiency.
\end{abstract}


\section{Introduction}
\IEEEPARstart{I}{mages} are often taken under sub-optimal lighting conditions, under the influence of backlit, uneven light, and dim light, due to inevitable environmental and/or technical constraints such as insufficient illumination and limited exposure time. Such images suffer from the compromised aesthetic quality and unsatisfactory transmission of information for high-level tasks such as object tracking, recognition, and detection\cite{li2021low}. Therefore, low-light image enhancement (LLIE) has attracted extensive attention from both academia and industry, aiming to recover hidden information and improve the quality of low-light images.

Over the past few years, prolific algorithms have been proposed to address this classic ill-posed problem, which can be roughly categorized into two groups: traditional methods (e.g., Histogram Equalization-based methods\cite{ibrahim2007brightness},\cite{abdullah2007dynamic}, and Retinex model-based methods\cite{guo2016lime},\cite{li2018structure},\cite{gu2019novel}) and deep learning-based methods\cite{lore2017llnet,ma2022toward,wei2018deep,jiang2021enlightengan,zhang2019kindling,zhang2021beyond,xu2022snr,liu2021retinex,yang2020fidelity}. The former formulates the degradation as a physical model and treats enhancement as the problem of estimating model parameters. Limited in characterizing diverse low-light factors, they mostly generate unsatisfying results and requires massive hand-crafted priors. The latter has carefully designed various models to adjust brightness, color tone and contrast, which can automatically learn from massive amounts of data. With the advanced design of deep neural networks, deep learning-based solutions enjoy better accuracy, robustness, and speed over conventional methods. However, most of them are based on the spatial information and rarely explore frequency domain information, which has been proved to be effective in improving the image quality\cite{fuoli2021fourier,jiang2021focal}.

Recently, some articles\cite{wang2023fourllie,li2023embedding,huang2022deep} have explored the Fourier frequency information of LLIE. They find that, in the Fourier space, the amplitude component of an image mainly reflects the lightness representation, while the phase component corresponds to structures or noise and is less related to lightness. Based on this observation, they integrate both the Fourier frequency and spatial information into neural networks and achieve impressive results. Despite the remarkable advancement, there are still several major issues. 

\textbf{(1) Inefficient utilization of frequency domain information.}
Although existing methods\cite{wang2023fourllie,li2023embedding,huang2022deep} have explored the application of frequency domain information in LLIE tasks, they have not sufficiently explored the potential of enhancement from the perspective of frequency domain. For example, \cite{wang2023fourllie} and \cite{huang2022deep} simply interact  the features in the spatial domain and the frequency domain by adding or weighted addition, while \cite{li2023embedding} adopts a concatenation operation, which limits the performance improvement. How to embed frequency domain prior information into spatial features and effectively incorporate them into a unified LLIE network is a vital issue.

\IEEEpubidadjcol

\textbf{(2) Inefficient information flow.}
The network structures of \cite{wang2023fourllie,li2023embedding,huang2022deep} are all in the form of multi-stage architecture. The basic idea is to decompose a challenging task into multiple sub-tasks and adopt progressive learning, which is conducive to performance improvement. However, in these network structures, the efficiency of information propagation and flow is low. For example,  \cite{wang2023fourllie,li2023embedding,huang2022deep} all use the output of the first stage directly as the input of the second stage, which ignores the propagation of intermediate features from the earlier to the later stages. Thus, the complementary information will be gradually weakened with the increasing network depths. In addition, they also ignore the exchange of cross-scale information of in-stage and cross-stage, and thereby fails to provide more precise and contextually enriched feature representations.

To address the above issues, we develop a novel Frequency-Spatial Interaction Driven Network (FSIDNet) for LLIE, which explores the efficient incorporation between spatial and frequency domains. The FSIDNet is a two-stage network, including an amplitude-guided brightness enhancement stage and a phase-guided structure-refined stage.  The first stage is designed to restore the amplitude of low-light images to improve the lightness, and the second stage restores phase information to refine fine-grained structures. In order to improve the complementary ability of frequency domain and spatial domain information, we carefully design two frequency-spatial interaction blocks (e.g., frequency-spatial interaction amplitude block for the first stage and frequency-spatial interaction phase block for the second stage) to mutually fuse frequency and spatial information to boost LLIE performance. Secondly, in order to effectively promote the propagation and flow of information in the two-stage network structure, we construct the Information Exchange Module (IEM) to associate two stages by adequately incorporating cross-stage and cross-scale features. In particular, a dynamic filter block is equipped in the IEM to generate dynamic filters based on the fused features, which can flexibly adapt to image contents.

Overall, the contributions of our work can be concluded as:

1) We develop a novel frequency-spatial interaction driven network (FSIDNet) for low-light image enhancement, consisting of an amplitude-guided brightness enhancement stage and a phase-guided structure-refined stage, which embraces the advantages of high performance and lower model complexity.

2) We carefully design two frequency-spatial interaction blocks to mutually fuse global frequency information and local spatial information. Thanks to its powerful modeling ability, the complementary information contained in spatial and frequency domains can be fully explored and utilized.

3) we carefully construct the Information Exchange Module (IEM) to associate two stages by adequately incorporating cross-stage and cross-scale features. This design effectively promotes the propagation and flow of information in the two-stage network.

4) We conduct experiments to verify the superiority of the proposed method. Experimental results on several widely used benchmark datasets (i.e., LOL-Real\cite{yang2021sparse}, LSRW-Huawei\cite{hai2023r2rnet}, etc) demonstrate that our method achieves the excellent performance in terms of visual results and quantitative metrics while preserving good model efficiency.

\section{Related work }
\subsection{Low-Light Image Enhancement}
Low-light image enhancement has received extensive attention from researchers as an important support for various downstream tasks\cite{xue2020arbitrarily,dong2023retrieving}. Traditional low-light image enhancement techniques mainly focus on constructing physical models using two types of methods, adaptive histogram equalisation\cite{pisano1998contrast} and Retinex theory\cite{land1971lightness}, which are processed by optimizing the parameter information of the image itself. The former class of algorithms optimizes pixel brightness based on the idea of histogram equalization, while the latter class of methods obtains the desired reflectance map (i.e.,the normal image) by estimating the light from the low-light input and removing the effect of the estimated light. For example, \cite{ibrahim2007brightness} partitions the smoothed histogram based on its local maximums, and then the histogram equalization process is applied independently to these partitions based on the new dynamic range. \cite{li2018structure} propose a robust Retinex model that additionally considers the noise map to improve the performance of enhancing low-light images with strong noise. Based on the robust Retinex model, an optimization function that includes novel regularization terms for the illumination and reflectance is also presented. \cite{gu2019novel}  propose a novel Retinex-based fractional-order variational model which decompose directly in the image domain and perform the fractional-order gradient total variation regularization on both the reflectance component and the illumination component to get more appropriate estimated results. 

The rapid development of deep learning has also triggered the enthusiasm of researchers to explore the field of low-light image enhancement. Numerous low-light enhancement algorithms through data-driven enhancement have been proposed one after another\cite{lore2017llnet,ma2022toward,wei2018deep,jiang2021enlightengan,zhang2019kindling,zhang2021beyond,xu2022snr,liu2021retinex,yang2020fidelity}. Lore et al.\cite{lore2017llnet} proposed LLNet, the first network that applies deep learning to image enhancement, which is trained on degraded images through an encoder-decoder architecture. \cite{wei2018deep} proposed Retinex-Net, which first introduced Retinex theory to deep learning and constructed an end-to-end image decomposition algorithm. Enlightengan\cite{jiang2021enlightengan} used a generative inverse network as the main framework and was first trained using unpaired images. Zhang et al.\cite{zhang2019kindling} proposed the KinD method to improve the problem of producing unnatural enhancement results in Retinex-Net by introducing training loss and adjusting the network architecture. \cite{xu2022snr}utilizes a signal-to-noise ratio aware transformer and a CNN model with spatially
varying operations for recovery. Although all these methods
have achieved remarkable results, they may pay less attention to the Fourier frequency information, which is effective for LLIE.

Recently, Fourier frequency information has opened up new ideas for deep-learning based LLIE and has achieved some developments. Huang et al.\cite{huang2022deep} pointed out amplitude component can reflect lightness representation of under-/over- exposure images. Li et al.\cite{li2023embedding} found lightness and noise can be decomposed in the Fourier space. Wang et al.\cite{wang2023fourllie} further explore the relationship between the amplitude components of low/normal-light images.

\subsection{Fourier Transform}
The Fourier transform is a widely used technique to analyze the frequency content in signals. It can be viewed as a global statistical information of signals, and thus can capture long-range dependency. Depend on the characteristic of Fourier transform, Fourier transform is used to perform computer vision tasks. Yang et al.\cite{yang2020fda} utilize Fourier transform to assist domain adaption for boosting cross domain semantic segmentation. Later on, the work of \cite{xu2021fourier} performs domain generation from Fourier-based perspective. A computationally efficient image classification network equipped with Fourier transform is introduced in \cite{rao2021global}. In addition, to improve perceptual quality and recover hard high-frequency details, the works of \cite{fuoli2021fourier,jiang2021focal} devise Fourier-based loss functions. In low-level tasks, Mao et al. \cite{mao2021deep} develop a Res FFTConv block to capture both long- and short- range dependencies for enhancing the details. Zhou et al. \cite{zhou2022spatial} propose to recover phase and amplitude seperatively with pan as guidance. Han et al. \cite{han2023dual} attempt to recover details from low-light images by selectively fusing high-frequency features at different levels and reconstruct high-frequency features back to low-frequency features in a certain way. Yu et al. \cite{yu2022frequency} build amplitude guided phase module to perform dehazing while Huang et al. \cite{huang2022deep} first recover amplitude and then recover phase to improve image lightness.

\section{Method}
The task of LLIE problem is to improve the perception or interpretability of an image captured in an environment with poor illumination. For a low-light image \textit{I} $\in\mathbb{R}^{W\times H\times 3}$ of width \textit{W} and height \textit{H}, the deep learning-based process can be modeled as 
\begin{equation}
\label{deqn_ex1a}
\hat{R} = \mathcal{F}(\textit{I}; \theta) ,
\end{equation}
where $\hat{R} \in\mathbb{R}^{W\times H\times 3}$ is the enhanced result and $\mathcal{F}$ represents the network with trainable parameters $\theta$. Currently, most existing deep learning-based LLIE methods separately restore various degradation in the spatial domain, ignoring the Fourier frequency domain information, which has been proved crucial for improving the image quality\cite{fuoli2021fourier}. To this end, we proposed a novel LLIE network termed FSDNet, based on frequency-spatial dual domain information and two-stage architecture.

\subsection{Fourier Frequency Information}
Firstly, we reviewed the operation and property of the Fourier transform. Given a single channel image \textit{x} with the shape of \textit{W} $\times$ \textit{H}, the Fourier transform $\mathcal{F}$ converts to the Fourier space as a complex component \textit{X}, which is expressed as:
\begin{footnotesize}
    \begin{equation}
    \label{deqn_ex1a}
    \mathcal{F}(x)(u,v) = \textit{X}(u,v)=\dfrac{1}{\sqrt{HW}}\sum_{h=0}^{H-1}\sum_{w=0}^{W-1}x(h,w)e^{-j2\pi(\dfrac{h}{H}u+\dfrac{w}{W}v)}
    \end{equation}
\end{footnotesize}

where \textit{h},\textit{w} are the coordinates in the spatial space and \textit{u},\textit{v} are the coordinates in the Fourier space, \textit{j} is the imaginary unit, the inverse process of $\mathcal{F}$ is denoted as $\mathcal{F}^{-1}$. Since an image or feature may contain multiple channels, we apply the Fourier transform separately to each channel in our work with the FFT\cite{prince1994fast}.
X(\textit{u},\textit{v}) consists of complex values
and can be represented by:
\begin{equation}
\label{deqn_ex1a}
X(u,v)=R(X(u,v))+jI(X(u,v)),
\end{equation}
where $R(X(u,v))$ and $I(X(u,v))$ are the real and imaginary parts of $X(u,v)$, respectively.

In the Fourier space, each complex component $X(u,v)$ can be represented by the amplitude component $\mathcal{A}(X(u,v))$ and the phase component $\mathcal{P}(X(u,v))$, which provides an intuitive analysis of the frequency components\cite{fuoli2021fourier}. These two components are expressed as:
\begin{equation}
\label{deqn_ex1a}
\begin{split}
    &\mathcal{A}(X(u,v))=\sqrt{R^2(X(u,v))+I^2(X(u,v))} \\
    &\mathcal{P}(X(u,v))=arctan[\dfrac{I(X(u,v))}{R(X(u,v))}]
\end{split}
\end{equation}
where $R(x)$ and $I(x)$ can also be obtained by:
\begin{equation}
\label{deqn_ex1a}
\begin{split}
    &\mathcal{R}(X(u,v))=\mathcal{A}(X(u,v))\times cos(\mathcal{P}(X(u,v)))\\
    &\mathcal{I}(X(u,v))=\mathcal{A}(X(u,v))\times sin(\mathcal{P}(X(u,v)))
\end{split}
\end{equation}

According to previous Fourier-based methods\cite{huang2022deep,wang2023fourllie}, we conclude that: in the Fourier space, the amplitude component of an image reflects the lightness representation, while the phase component corresponds to the structures and noises.This inspires us to process lightness and structures separately in the Fourier domain. In this way, we designed a novel two-stage network for LLIE by simultaneously processing the spatial-frequency dual domain information.

\subsection{The FSIDNet Network}
The entire network, as shown in Figure 2, contains an amplitude-guided brightness enhancement stage and a phase-guided structure-refined stage. The former restores the amplitude of low-light images to improve the lightness, and the latter restores the phase information to refine fine-grained structures. The detailed implementations are given in the following parts. Since each stage is designed by a task-specific subnetwork, both the input and supervision signals should be different.

Let $X_{in}$ and $X_{gt}$ denote low-light images and ground truth images, $Y_1$ and $Y_2$ represent the outputs of two stages. Since each stage is designed by a task-specific subnetwork, both the input and supervision signals should be different. For the first stage, the low-light images $X_{in}$ are served as input, $\mathcal{F}^{-1}(\mathcal{A}(X_{gt}),\mathcal{P}(X_{in}))$ and $\mathcal{A}(X_{gt})$ are used to supervise the learning of amplitude representation. For the second stage, the $X_{gt}$ are used as the supervision signal, but we do not directly use $Y_1$ as the input of the second stage. Instead, we use $\mathcal{F}^{-1}(\mathcal{A}(Y_1),\mathcal{P}(X_{in}))$ to ensure the retention of original phase information.

To achieve an efficient target, we design an encoder-decoder-like (or U-Net-like)\cite{ronneberger2015u} format for each stage, consisting of seven basic units ( e.g., frequency-spatial interaction amplitude block (FSIA) and frequency-spatial interaction phase block (FSIP)), three downsampling blocks and three upsampling blocks in each stage. In the first stage, except for the lowest scale, each scale contains a skip connection between the encoder and decoder. However, considering that the decoder of the second stage integrates features from the previous stage and the current encoder by adopting the proposed fusion module, we do not take a similar approach.

To further facilitate learning the amplitude and phase representations, we propose the FSIA block and FSIP block as the basic unit of the two sub-networks, respectively. We illustrate the FSIA block as shown in Figure 3. Specifically, it comprises a spatial branch and a frequency branch for processing spatial and frequency representations. Denoting $f_i$ as the input features of FSIA block, the spatial branch first adopts a residual block with 3 × 3 convolution layers to process information in the spatial domain and obtain $f_{s1}$. While the frequency branch uses a 1 × 1 convolution to process $f_i$ first that obtains $f_{f0}$, and then adopts Fourier transform to convert it to the Fourier space as $F_{f0}$ by Eq. 2. To process frequency-domain representation $F_{f0}$, we adopt the operation $Op(·)$ that consists of 1 × 1 convolution layers on its amplitude component, and then recompose the operated result with the phase component that obtain $f_{f1}$, which is expressed as:
\begin{equation}
\label{deqn_ex1a}
f_{f1}=\mathcal{F}^{-1}(Op(\mathcal{A}(F_{f0})), \mathcal{P}(F_{f0})).
\end{equation}
Thus, $f_{f1}$ is the processed result of the frequency-domain representation. Next, we interact the features from spatial branch $f_{s1}$ and frequency branch $f_{f1}$ as:
\begin{equation}
\label{deqn_ex1a}
\begin{split}
    f^{'}_{s1}=f_{s1}+W_{1}(f_{f1}),\\
    f^{'}_{f1}=f_{f1}+W_{2}(f_{s1}),
\end{split}
\end{equation}
where both $W_1(\cdot)$ and $W_2(\cdot)$ denote the 3 × 3 convolution operation, $f^{'}_{s1}$ and $f^{'}_{f1}$ are the output of the interacted spatial branch and frequency branch. In that case, both $f^{'}_{s1}$ and $f^{'}_{f1}$ get the complementary representation, which benefits for these two branches to obtain more representational features. The following spatial and frequency branches are formulated in the same way as above and output the results $f_{s2}$ and $f_{f2}$, respectively.

\begin{figure*}[ht] 
    \centering 
    \includegraphics[scale=0.8]{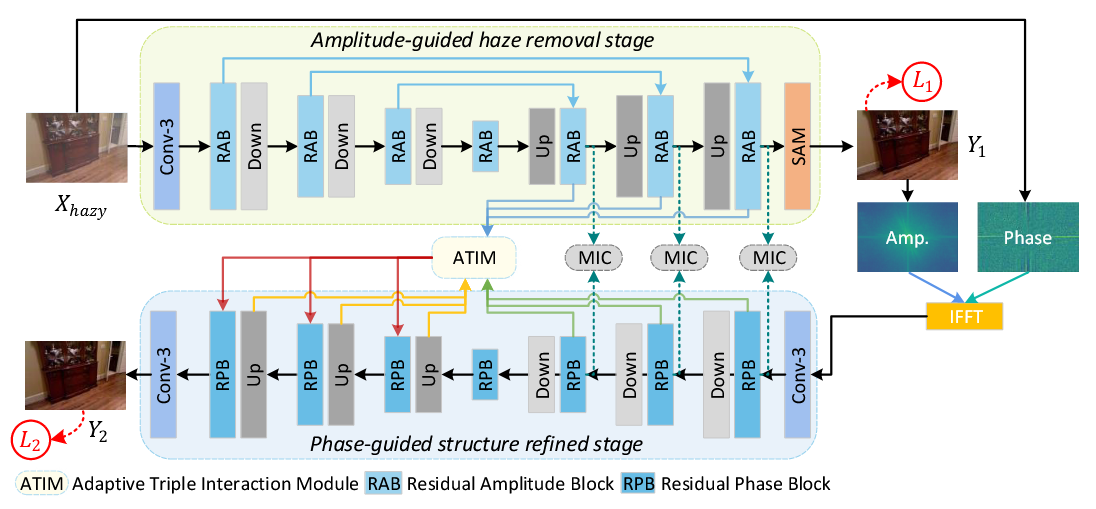} 
    \caption{The proposed FSIDNet Network.}
    \label{fig5} 
\end{figure*}

Finally, we concatenate $f_{s2}$ and $f_{f2}$ and then apply a 1 × 1 convolution operation to integrate them as $f_o$, which is the output of FSIA block. Similarly, for the FSIP block, we replace the operation on the amplitude component in Eq. 6 with the phase component, while other parts keep unchanged.

\subsection{Information Exchange Module}
The information exchange is an important ingredient in the two-stage design because simply passing the previous stage’s output to the next stage neglects the potential intermediate features, but the original, useful shallow features will be gradually weakened with the increasing network depths. In our framework, we construct the Information Exchange Module (IEM) to associate two stages by adequately incorporating cross-scale and cross-stage features. The detailed implementations are in the following parts.

\textbf{The cross-scale interaction.} In the first stage’s decoder and second stage’s encoder, the IEM integrates information from all scales, allowing for both top-down and bottom-up information flow. Further, this procedure combines features with different receptive fields, which enriches contextual representations.  As shown in Figure 4, the features $\{D_{1}^{si}\} ^{3}_{i=1}$ in the decoder of the first stage firstly will be extracted by three independent 3 × 3 convolutions. Upsampling and downsampling operations will further resize the obtained resolution features. To achieve the combination of precisely spatial high-resolution features with low-resolution features containing rich contextual information, a simple channel concatenation operation followed by 1 × 1 convolution is used to fuse them. The resultant multi-scale features can be denoted as  $\{\hat{D}_{1}^{si}\} ^{3}_{i=1}$. Similarly, we can obtain the fused features $\{\hat{E}_{2}^{si}\} ^{3}_{i=1}$ in the encoder of the second stage.

\textbf{The cross-stage interaction.} The IEM is equipped with an dynamic filter block(DFB), which generates weight filters with flexibility based on the fused cross-domain and cross-scale features and applies them to the decoder’s features of the second stage for representation capability enhancement. The DFB first uses one 1 × 1 convolution to fuse three inputs and then embeds resulting features into lightweight spatial and channel context branches to enhance filter representations. The spatial context branch includes a 3 × 3 depthwise convolution, whereas the channel context branch consists of a pooling layer and a convolutional layer. To take full advantage of them, the element-wise addition operation is used to integrate them. In order to formulate spatially-varying filters, a 1 × 1 convolutional layer is utilized to produce feature maps, and then reshape them into a series of per-pixel kernels. In this way, a series of content-adaptive filters for each location is learned. Finally, taking the specific-scale feature $D^{s1}_{2}$ as an example, the dynamic filter results can be written as:
\begin{equation}
\label{deqn_ex1a}
\hat{D}^{s1}_{2}=D^{s1}_{2}*W+D^{s1}_{2},
\end{equation}
where “*” indicates the convolution operation, $\hat{D}^{s1}_{2}$ is the enhanced features in the decoder of the second stage. Here, we adopt residual learning to preserve more low-level features.

\begin{table*}[!t]
\caption{Quantitative comparison on the LOL-Real\cite{yang2021sparse}, LOL-Synthetic\cite{yang2021sparse}, LSRW-Huawei\cite{hai2023r2rnet}, and LSRW-Nikon\cite{hai2023r2rnet}. The best results are highlighted in red and the second-best ones are marked in blue.\label{tab:table1}}
\centering
\begin{tabular}{ccccccccc}
\hline
\multirow{2}{*}{Methods}&\multicolumn{2}{c}{LOL-Real\cite{yang2021sparse}}&\multicolumn{2}{c}{LOL-Synthetic\cite{yang2021sparse}}&\multicolumn{2}{c}{LSRW-Huawei\cite{hai2023r2rnet}}&\multicolumn{2}{c}{LSRW-Nikon\cite{hai2023r2rnet}}\\
\cmidrule(r){2-3} \cmidrule(r){4-5} \cmidrule(r){6-7} \cmidrule(r){8-9}
&PSNR\textbf{$\uparrow$}&SSIM\textbf{$\uparrow$}&PSNR\textbf{$\uparrow$}&SSIM\textbf{$\uparrow$}&PSNR\textbf{$\uparrow$}&SSIM\textbf{$\uparrow$}&PSNR\textbf{$\uparrow$}&SSIM\textbf{$\uparrow$}\\
\midrule
LIME\cite{guo2016lime} & 15.24 & 0.4190 &16.88 &0.7578&17.00&0.3816&13.53&0.3321\\
MF\cite{fu2016fusion} & 18.72 & 0.5089& 17.50 & 0.7737& 18.26& 0.4283& 15.44&0.3997\\
NPE\cite{wang2013naturalness} & 17.33 & 0.4642& 16.60& 0.7781& 17.08& 0.3905& 14.86& 0.3738\\
SRIE\cite{fu2016weighted} & 14.45 & 0.5240 & 14.50& 0.6640& 13.42& 0.4282& 13.26& 0.3963\\
DRD\cite{wei2018deep} &16.08 &0.6555 &18.28& 0.7737&18.23&0.5220&15.18&0.3809\\
Kind\cite{zhang2019kindling} &20.01 & 0.8412 & 22.62& 0.9041& 16.58& 0.5690& 11.52&0.3827\\
Kind++\cite{zhang2021beyond} &20.59&0.8294& 21.17&0.8814&15.43&0.5695&14.79&0.4749\\
MIRNet\cite{zamir2020learning} &22.11 & 0.7942& 22.52& 0.8997& 19.98&0.6085&17.10&0.5022\\
SGM\cite{yang2021sparse} &20.06 &0.8158&22.05&0.9089&18.85&0.5991&15.73&0.4971\\
HDMNet\cite{liang2022learning} &18.55&0.7132&20.54&0.8539&20.81&0.6071&16.65&0.4870\\
SNR-Aware\cite{xu2022snr} &21.48&0.8478&24.13&0.9269&20.67&0.5910&17.54&0.4822\\
FourLLIE\cite{wang2023fourllie} &22.34&0.8468&24.65&0.9192&21.30&0.6220&17.82&0.5036\\
\hline
OURS&22.65&0.8524&24.93&0.9187&21.71&0.6184&17.80&0.5069\\
\end{tabular}
\end{table*}

\subsection{Loss Function}
Since there are two sub-networks in our FSDNet, then the loss functions of these two sub-networks are expressed respectively as:
\begin{equation}
\label{deqn_ex1a}
L_{s1}=\vert \vert Y_1-\mathcal{F}^{-1}(\mathcal{A}(X_{gt}),\mathcal{P}(X_{in}))\vert \vert_{1}+\alpha\vert \vert \mathcal{A}(Y_{1})-\mathcal{A}(X_{gt})\vert \vert_{1},
\end{equation}

\begin{equation}
\label{deqn_ex1a}
L_{s2}=\vert \vert Y_2-X_{gt}\vert \vert_{1}+\beta\vert \vert \mathcal{P}(Y_{2})-\mathcal{P}(X_{gt})\vert \vert_{1},
\end{equation}
where $\vert\vert\cdot\vert\vert$ denotes the Mean Absolute Error (MAE), $\alpha$ and $\beta$ are the trade-off factor, and we set them as 0.2, 0.1, respectively. Both terms in each equation above are performed on spatial and frequency domains, respectively. The overall network comprised is training in an end-to-end manner, and the overall loss $L_{total}$ is the combination of $L_{s1}$ and $L_{s2}$, which is formulated as:
\begin{equation}
\label{deqn_ex1a}
L_{total}=L_{s2}+\lambda L_{s1},
\end{equation}
where $\lambda$ is the weight factor and is empirically set as 0.5. In this way, the FSDNet is able to conduct low-light image enhancement in a coarse to fine manner as shown in Fig. 2.

\section{Experiment}
\subsection{Experiment Setting}
\textbf{Datasets}. For fair comparisons with existing LLIE methods, we choose four widely used LLIE datasets for evaluating FSDNet, including LOL-Real\cite{yang2021sparse}, LOL-Synthetic\cite{yang2021sparse}, LSRW-Huawei\cite{hai2023r2rnet}, and LSRW-Nikon\cite{hai2023r2rnet}. LOL-Real is captured in real scenes by changing exposure time and ISO. It contains 689 low-/normallight image pairs for training and 100 low-/normal- light image pairs for testing. It is worth noting that LOL-Real is the extended version of LOL\cite{wei2018deep}, which contains all the image pairs of LOL, so we only evaluate in LOL-Real. LOL-Synthetic is synthesized from raw images by analyzing the distribution of luminance channels of low-/normallight images. It contains 900 low-/normal- light image pairs for training and 100 low-/normal- light image pairs for testing. LSRW-Huawei and LSRW-Nikon are captured in real scenes like LOL-Real but with different devices. LSRW-Huawei is collected by a Huawei P40 Pro and LSRW-Nikon is collected by a Nikon D7500. LSRW-Huawei contains 3150 training image pairs and 20 testing image pairs. LSRW-Nikon contains 2450 training image pairs and 30 testing image pairs. Besides, we also evaluate FSDNet on three unpaired datasets DICM\cite{lee2013contrast}, LIME\cite{guo2016lime}, MEF\cite{ma2015perceptual}.

\textbf{Implementation Details.}
The implement of our proposed method is based on PyTorch framework with one NVIDIA 3090 GPU. During the training, we adopt the ADAM optimizer with $\beta_{1}=0.9$, $\beta_{2}=0.999$, $\epsilon=10^{-8}$. We set the initial learning rate as 0.0002 and then linearly decrease it to half every 100 epoch. In addition, the input images are cropped to 384 × 384 randomly for each training. And in each mini-batch, we enrich and augment these patches to expand training samples by flipping horizontally or vertically and rotating 90°.

\subsection{Performance Evaluation}
In this paper, we compare our algorithm with several state-of-the-art LLIE methods, including traditional methods LIME\cite{guo2016lime}, MF\cite{fu2016fusion}, NPE\cite{wang2013naturalness}, SRIE\cite{fu2016weighted} and deep learning-based methods DRD\cite{wei2018deep}, Kind\cite{zhang2019kindling}, Kind++\cite{zhang2021beyond}, MIRNet\cite{zamir2020learning}, SGM\cite{yang2021sparse}, HDMNet\cite{liang2022learning}, SNR-Aware\cite{xu2022snr}, FourLLIE\cite{wang2023fourllie}. Note that all deep learning-based methods are trained on the same datasets with respective public codes.

\textbf{Quantitative comparison.} In this section, the proposed algorithms are compared quantitatively and experimentally using three metrics: peak signal-to-noise ratio (PSNR)\cite{huynh2008scope}, structural similarity (SSIM)\cite{wang2004image}, natural image quality evaluator(NIQE)\cite{mittal2012making}. PSNR and SSIM are full-reference image quality assessment methods that indicate the noise level and structural similarity between the enhancement result and the reference image, respectively. Generally, the higher PSNR, higher SSIM indicate better image recovery quality. NIQE is a non-reference image quality assessment method that evaluates the naturalness of the image, and a lower value indicates better quality. 

\begin{table}
\caption{NIQE scores on on three unpaired datasets. THE BEST RESULTS ARE
HIGHLIGHTED IN RED AND THE SECOND-BEST ONES ARE MARKED IN BLUE}\label{datasettable}
\centering
\begin{tabular}{ccccc}
\toprule
\textbf{Methods} & \textbf{LIME} & \textbf{DICM} & \textbf{MEF} & \textbf{AVG} \\
\midrule
Kind & 4.772 & 3.614 & 4.819 & 4.4017 \\
MIRNet  & 6.453 & 4.042 & 5.504 & 5.3330 \\
SGM  & 5.451 & 4.733 & 5.754 & 5.3127 \\
HDMNet & 6.403 & 4.773 & 5.993 & 5.7230\\
SNR-Aware & 4.618 & 3.227 & 4.589 & 4.1447\\
FourLLIE & 4.402 &3.374 & 4.362 & 4.0460\\
\midrule
ours & 4.389 & 3.327 & 4.413 & 4.0430\\
\bottomrule
\end{tabular}
\end{table}

As shown in TABLE 1, our method achieves the best performance among these methods in most cases, for the rest it almost achieves the second-best. It should be mentioned that although there is no performance improvement on the LOL-Synthetic dataset compared to SNR-Aware\cite{xu2022snr}, our method is lighter and more efficient.

In addition, we measure the naturalness image quality evaluator (NIQE) score on three unpaired datasets. The images with the lower NIQE scores represent the higher naturalness image quality. Note that “AVG” represents the average values of the NIQE score on three datasets, and all methods are pre-trained on LSRW-Huawei. Table 2 presents the results of the NIQE evaluation. It can be seen that the proposed method outperforms most existing LLIE methods. Although SNR-Aware and FourLLIE reach competitive performance in the NIQE evaluation, which have much more complex network structure than the proposed method.

\textbf{Qualitative Comparison.} We also show the visualized comparisons with representative methods of the LOL datasets (Real+Synthetic) in Figure 4, and the visualized comparisons of the LSRW datasets (Huawei+Nikon) in Figure 5, respectively. It can be seen that the compared methods cause poor visual results, such as color distortion, incorrect lightness, loss of detail texture, etc. In contrast, our FSDNet preserves more details and involves fewer color distortions, resulting in the closest match to the ground-truth images and accurate lightness improvement.

Besides, we also present the visual comparison on unpaired datasets. Fig. 6 presents the results of DICM (first row) and MEF (second row).

\textbf{In-the-Wild Experimental Evaluation.} Low-light image enhancement in the wild scenarios is extremely challenging. The control of the partial overexposure information of the image, the correction of the overall color, and the preservation of image details are all problems that need to be solved urgently. Here, we tested lots of challenging in-the-wild examples from the DARK FACE\cite{yang2020advancing} and ExDark\cite{loh2019getting} datasets. As demonstrated in Fig. 7, through a large number of experiments, it can be seen that our method outperforms most existing LLIE methods.

\subsection{Ablation studies}
In this section, we further conduct experiments to verify the effectiveness of key components in FSIDNet.

\textbf{The effectiveness of the FSIA and FSIP:} First, we remove the frequency branch and only preserve the spatial branch in FSIA and FSIP block, and the remaining model is called MODEL1. Then, we remove the spatial branch and only preserve the frequency branch in FSIA and FSIP block, and the remaining model is called MODEL2. Finally, we disconnect the feature interaction between the spatial branch and the frequency branch (i.e. Formula 7 does not work), named MODEL3. The results are reported in Table 3. All quantitative metrics of MODEL3 are  better than those of MODEL1 and MODEL2, demonstrating that both frequency domain and spatial domain information are conducive to enhancing the representation ability of the model. However, the improvement is limited due to that concatenation is too simple to perform the interaction between the frequency domain and the spatial domain. Finally, equipped with our carefully designed FSIA and FSIP block, our method FSIDNet achieves the best performance in terms of both PSNR and SSIM, which demonstrates the effectiveness of integrating these two complementary representations.

\begin{table}
\caption{Ablation study of the proposed FSIA and FSIP}\label{datasettable}
\centering
\begin{tabular}{ccccc}
\toprule
\multirow{2}{*}{Model}&\multicolumn{2}{c}{LOL-Real\cite{yang2021sparse}}&\multicolumn{2}{c}{LSRW-Huawei\cite{hai2023r2rnet}}\\
\cmidrule(r){2-3} \cmidrule(r){4-5}&PSNR\textbf{$\uparrow$}&SSIM\textbf{$\uparrow$}&PSNR\textbf{$\uparrow$}&SSIM\textbf{$\uparrow$}\\
\midrule
MODEL1 & 21.52 & 0.8343 & 21.16 & 0.5984 \\
MODEL2  & 21.67 & 0.8380 & 20.97 & 0.5964 \\
MODEL3  & 22.18 & 0.8421 & 21.35 & 0.6082 \\
\midrule
FSIDNet & 22.65 & 0.8524 & 21.71 & 0.6184\\
\bottomrule
\end{tabular}
\end{table}

\begin{table}
\caption{Ablation study of the proposed IEM}\label{datasettable}
\centering
\begin{tabular}{ccccccc}
\toprule
\multirow{2}{*}{Model}&\multirow{2}{*}{Cross}&\multirow{2}{*}{Cross}&\multicolumn{2}{c}{LOL-Real\cite{yang2021sparse}}&\multicolumn{2}{c}{LSRW-Huawei\cite{hai2023r2rnet}}\\
\cmidrule(r){4-5} \cmidrule(r){6-7}&stage&scale&PSNR\textbf{$\uparrow$}&SSIM\textbf{$\uparrow$}&PSNR\textbf{$\uparrow$}&SSIM\textbf{$\uparrow$}\\
\midrule
MODEL4 & $\times$& $\times$& 22.13 & 0.8243 & 20.64 & 0.5837 \\
MODEL5  &$\checkmark$ &$\times$& 22.42 & 0.8365 & 20.98 & 0.6071 \\
MODEL6  &$\times$ &$\checkmark$ & 22.39 & 0.8481 & 21.02 & 0.5943 \\
\midrule
FSIDNet &$\checkmark$&$\checkmark$& 22.65 & 0.8524 & 21.71&0.6184\\
\bottomrule
\end{tabular}
\end{table}

\textbf{The effectiveness of the IEM:} We also conduct experiments to analyze the effectiveness of the IEM block. Since IEM consists of two parts, we analyze the impact of these two parts on performance separately. Specifically, we compare the results of the model with and without the IEM, the specific experimental configuration and results are shown in Table 4. From the aspects of quantitative metrics, the introduction of cross-stage interaction and cross-scale interaction can obviously improve the PSNR and SSIM performance of the model. Further, when these two parts are employed, the gains of PSNR and SSIM are larger than when these two parts work alone, which suggests utilizing the interactive cross-scale and cross-stage features is beneficial to improving the representation capability.

\section{Conclusion}

This paper proposes a novel Frequency-Spatial Interaction Driven Network (FSIDNet) for low-light image enhancement (LLIE) to address the limitations of existing methods that either ignore frequency domain information or fail to promote effective information propagation. FSIDNet adopts a two-stage architecture: the first stage restores the amplitude of low-light images to improve lightness, and the second stage restores phase information to refine fine-grained structures. Its key contributions include: 1) designing two frequency-spatial interaction blocks to mutually amalgamate complementary spatial and frequency information, fully exploring their potential for LLIE; 2) constructing an Information Exchange Module (IEM) that incorporates cross-stage and cross-scale features to effectively boost information propagation in the two-stage structure; and 3) achieving excellent performance in both visual results and quantitative metrics (on benchmark datasets like LOL-Real and LSRWHuawei) while preserving good model efficiency, as demonstrated by extensive experiments.


\bibliographystyle{IEEEtran}
\bibliography{references}

\begin{thebibliography}{10}
\providecommand{\url}[1]{#1}
\csname url@samestyle\endcsname
\providecommand{\newblock}{\relax}
\providecommand{\bibinfo}[2]{#2}
\providecommand{\BIBentrySTDinterwordspacing}{\spaceskip=0pt\relax}
\providecommand{\BIBentryALTinterwordstretchfactor}{4}
\providecommand{\BIBentryALTinterwordspacing}{\spaceskip=\fontdimen2\font plus
\BIBentryALTinterwordstretchfactor\fontdimen3\font minus \fontdimen4\font\relax}
\providecommand{\BIBforeignlanguage}[2]{{%
\expandafter\ifx\csname l@#1\endcsname\relax
\typeout{** WARNING: IEEEtran.bst: No hyphenation pattern has been}%
\typeout{** loaded for the language `#1'. Using the pattern for}%
\typeout{** the default language instead.}%
\else
\language=\csname l@#1\endcsname
\fi
#2}}
\providecommand{\BIBdecl}{\relax}
\BIBdecl

\bibitem{li2021low}
C.~Li, C.~Guo, L.~Han, J.~Jiang, M.-M. Cheng, J.~Gu, and C.~C. Loy, ``Low-light image and video enhancement using deep learning: A survey,'' \emph{IEEE transactions on pattern analysis and machine intelligence}, vol.~44, no.~12, pp. 9396--9416, 2021.

\bibitem{ibrahim2007brightness}
H.~Ibrahim and N.~S.~P. Kong, ``Brightness preserving dynamic histogram equalization for image contrast enhancement,'' \emph{IEEE Transactions on Consumer Electronics}, vol.~53, no.~4, pp. 1752--1758, 2007.

\bibitem{abdullah2007dynamic}
M.~Abdullah-Al-Wadud, M.~H. Kabir, M.~A.~A. Dewan, and O.~Chae, ``A dynamic histogram equalization for image contrast enhancement,'' \emph{IEEE transactions on consumer electronics}, vol.~53, no.~2, pp. 593--600, 2007.

\bibitem{guo2016lime}
X.~Guo, Y.~Li, and H.~Ling, ``Lime: Low-light image enhancement via illumination map estimation,'' \emph{IEEE Transactions on image processing}, vol.~26, no.~2, pp. 982--993, 2016.

\bibitem{li2018structure}
M.~Li, J.~Liu, W.~Yang, X.~Sun, and Z.~Guo, ``Structure-revealing low-light image enhancement via robust retinex model,'' \emph{IEEE Transactions on Image Processing}, vol.~27, no.~6, pp. 2828--2841, 2018.

\bibitem{gu2019novel}
Z.~Gu, F.~Li, F.~Fang, and G.~Zhang, ``A novel retinex-based fractional-order variational model for images with severely low light,'' \emph{IEEE Transactions on Image Processing}, vol.~29, pp. 3239--3253, 2019.

\bibitem{lore2017llnet}
K.~G. Lore, A.~Akintayo, and S.~Sarkar, ``Llnet: A deep autoencoder approach to natural low-light image enhancement,'' \emph{Pattern Recognition}, vol.~61, pp. 650--662, 2017.

\bibitem{ma2022toward}
L.~Ma, T.~Ma, R.~Liu, X.~Fan, and Z.~Luo, ``Toward fast, flexible, and robust low-light image enhancement,'' in \emph{Proceedings of the IEEE/CVF conference on computer vision and pattern recognition}, 2022, pp. 5637--5646.

\bibitem{wei2018deep}
C.~Wei, W.~Wang, W.~Yang, and J.~Liu, ``Deep retinex decomposition for low-light enhancement,'' \emph{arXiv preprint arXiv:1808.04560}, 2018.

\bibitem{jiang2021enlightengan}
Y.~Jiang, X.~Gong, D.~Liu, Y.~Cheng, C.~Fang, X.~Shen, J.~Yang, P.~Zhou, and Z.~Wang, ``Enlightengan: Deep light enhancement without paired supervision,'' \emph{IEEE transactions on image processing}, vol.~30, pp. 2340--2349, 2021.

\bibitem{zhang2019kindling}
Y.~Zhang, J.~Zhang, and X.~Guo, ``Kindling the darkness: A practical low-light image enhancer,'' in \emph{Proceedings of the 27th ACM international conference on multimedia}, 2019, pp. 1632--1640.

\bibitem{zhang2021beyond}
Y.~Zhang, X.~Guo, J.~Ma, W.~Liu, and J.~Zhang, ``Beyond brightening low-light images,'' \emph{International Journal of Computer Vision}, vol. 129, pp. 1013--1037, 2021.

\bibitem{xu2022snr}
X.~Xu, R.~Wang, C.-W. Fu, and J.~Jia, ``Snr-aware low-light image enhancement,'' in \emph{Proceedings of the IEEE/CVF conference on computer vision and pattern recognition}, 2022, pp. 17\,714--17\,724.

\bibitem{liu2021retinex}
R.~Liu, L.~Ma, J.~Zhang, X.~Fan, and Z.~Luo, ``Retinex-inspired unrolling with cooperative prior architecture search for low-light image enhancement,'' in \emph{Proceedings of the IEEE/CVF conference on computer vision and pattern recognition}, 2021, pp. 10\,561--10\,570.

\bibitem{yang2020fidelity}
W.~Yang, S.~Wang, Y.~Fang, Y.~Wang, and J.~Liu, ``From fidelity to perceptual quality: A semi-supervised approach for low-light image enhancement,'' in \emph{Proceedings of the IEEE/CVF conference on computer vision and pattern recognition}, 2020, pp. 3063--3072.

\bibitem{fuoli2021fourier}
D.~Fuoli, L.~Van~Gool, and R.~Timofte, ``Fourier space losses for efficient perceptual image super-resolution,'' in \emph{Proceedings of the IEEE/CVF International Conference on Computer Vision}, 2021, pp. 2360--2369.

\bibitem{jiang2021focal}
L.~Jiang, B.~Dai, W.~Wu, and C.~C. Loy, ``Focal frequency loss for image reconstruction and synthesis,'' in \emph{Proceedings of the IEEE/CVF international conference on computer vision}, 2021, pp. 13\,919--13\,929.

\bibitem{wang2023fourllie}
C.~Wang, H.~Wu, and Z.~Jin, ``Fourllie: Boosting low-light image enhancement by fourier frequency information,'' in \emph{Proceedings of the 31st ACM International Conference on Multimedia}, 2023, pp. 7459--7469.

\bibitem{li2023embedding}
C.~Li, C.-L. Guo, M.~Zhou, Z.~Liang, S.~Zhou, R.~Feng, and C.~C. Loy, ``Embedding fourier for ultra-high-definition low-light image enhancement,'' \emph{arXiv preprint arXiv:2302.11831}, 2023.

\bibitem{huang2022deep}
J.~Huang, Y.~Liu, F.~Zhao, K.~Yan, J.~Zhang, Y.~Huang, M.~Zhou, and Z.~Xiong, ``Deep fourier-based exposure correction network with spatial-frequency interaction,'' in \emph{European Conference on Computer Vision}.\hskip 1em plus 0.5em minus 0.4em\relax Springer, 2022, pp. 163--180.

\bibitem{yang2021sparse}
W.~Yang, W.~Wang, H.~Huang, S.~Wang, and J.~Liu, ``Sparse gradient regularized deep retinex network for robust low-light image enhancement,'' \emph{IEEE Transactions on Image Processing}, vol.~30, pp. 2072--2086, 2021.

\bibitem{hai2023r2rnet}
J.~Hai, Z.~Xuan, R.~Yang, Y.~Hao, F.~Zou, F.~Lin, and S.~Han, ``R2rnet: Low-light image enhancement via real-low to real-normal network,'' \emph{Journal of Visual Communication and Image Representation}, vol.~90, p. 103712, 2023.

\bibitem{xue2020arbitrarily}
M.~Xue, P.~Shivakumara, C.~Zhang, Y.~Xiao, T.~Lu, U.~Pal, D.~Lopresti, and Z.~Yang, ``Arbitrarily-oriented text detection in low light natural scene images,'' \emph{IEEE Transactions on Multimedia}, vol.~23, pp. 2706--2720, 2020.

\bibitem{dong2023retrieving}
K.~Dong, Y.~Guo, R.~Yang, Y.~Cheng, J.~Suo, and Q.~Dai, ``Retrieving object motions from coded shutter snapshot in dark environment,'' \emph{IEEE Transactions on Image Processing}, vol.~32, pp. 3281--3294, 2023.

\bibitem{pisano1998contrast}
E.~D. Pisano, S.~Zong, B.~M. Hemminger, M.~DeLuca, R.~E. Johnston, K.~Muller, M.~P. Braeuning, and S.~M. Pizer, ``Contrast limited adaptive histogram equalization image processing to improve the detection of simulated spiculations in dense mammograms,'' \emph{Journal of Digital imaging}, vol.~11, pp. 193--200, 1998.

\bibitem{land1971lightness}
E.~H. Land and J.~J. McCann, ``Lightness and retinex theory,'' \emph{Josa}, vol.~61, no.~1, pp. 1--11, 1971.

\bibitem{yang2020fda}
Y.~Yang and S.~Soatto, ``Fda: Fourier domain adaptation for semantic segmentation,'' in \emph{Proceedings of the IEEE/CVF conference on computer vision and pattern recognition}, 2020, pp. 4085--4095.

\bibitem{xu2021fourier}
Q.~Xu, R.~Zhang, Y.~Zhang, Y.~Wang, and Q.~Tian, ``A fourier-based framework for domain generalization,'' in \emph{Proceedings of the IEEE/CVF conference on computer vision and pattern recognition}, 2021, pp. 14\,383--14\,392.

\bibitem{rao2021global}
Y.~Rao, W.~Zhao, Z.~Zhu, J.~Lu, and J.~Zhou, ``Global filter networks for image classification,'' \emph{Advances in neural information processing systems}, vol.~34, pp. 980--993, 2021.

\bibitem{mao2021deep}
X.~Mao, Y.~Liu, W.~Shen, Q.~Li, and Y.~Wang, ``Deep residual fourier transformation for single image deblurring. arxiv 2021,'' \emph{arXiv preprint arXiv:2111.11745}, vol.~2, no.~3, p.~6, 2021.

\bibitem{zhou2022spatial}
M.~Zhou, J.~Huang, K.~Yan, H.~Yu, X.~Fu, A.~Liu, X.~Wei, and F.~Zhao, ``Spatial-frequency domain information integration for pan-sharpening,'' in \emph{European conference on computer vision}.\hskip 1em plus 0.5em minus 0.4em\relax Springer, 2022, pp. 274--291.

\bibitem{han2023dual}
G.~Han, K.~Wu, F.~Zeng, J.~Liu, and S.~Kwong, ``Dual-stream adaptive convergent low-light image enhancement network based on frequency perception,'' \emph{IEEE Transactions on Computational Imaging}, 2023.

\bibitem{yu2022frequency}
H.~Yu, N.~Zheng, M.~Zhou, J.~Huang, Z.~Xiao, and F.~Zhao, ``Frequency and spatial dual guidance for image dehazing,'' in \emph{European Conference on Computer Vision}.\hskip 1em plus 0.5em minus 0.4em\relax Springer, 2022, pp. 181--198.

\bibitem{prince1994fast}
E.~Prince and E.~Prince, ``The fast fourier transform,'' \emph{Mathematical Techniques in Crystallography and Materials Science}, pp. 140--156, 1994.

\bibitem{ronneberger2015u}
O.~Ronneberger, P.~Fischer, and T.~Brox, ``U-net: Convolutional networks for biomedical image segmentation,'' in \emph{Medical image computing and computer-assisted intervention--MICCAI 2015: 18th international conference, Munich, Germany, October 5-9, 2015, proceedings, part III 18}.\hskip 1em plus 0.5em minus 0.4em\relax Springer, 2015, pp. 234--241.

\bibitem{fu2016fusion}
X.~Fu, D.~Zeng, Y.~Huang, Y.~Liao, X.~Ding, and J.~Paisley, ``A fusion-based enhancing method for weakly illuminated images,'' \emph{Signal Processing}, vol. 129, pp. 82--96, 2016.

\bibitem{wang2013naturalness}
S.~Wang, J.~Zheng, H.-M. Hu, and B.~Li, ``Naturalness preserved enhancement algorithm for non-uniform illumination images,'' \emph{IEEE transactions on image processing}, vol.~22, no.~9, pp. 3538--3548, 2013.

\bibitem{fu2016weighted}
X.~Fu, D.~Zeng, Y.~Huang, X.-P. Zhang, and X.~Ding, ``A weighted variational model for simultaneous reflectance and illumination estimation,'' in \emph{Proceedings of the IEEE conference on computer vision and pattern recognition}, 2016, pp. 2782--2790.

\bibitem{zamir2020learning}
S.~W. Zamir, A.~Arora, S.~Khan, M.~Hayat, F.~S. Khan, M.-H. Yang, and L.~Shao, ``Learning enriched features for real image restoration and enhancement,'' in \emph{Computer Vision--ECCV 2020: 16th European Conference, Glasgow, UK, August 23--28, 2020, Proceedings, Part XXV 16}.\hskip 1em plus 0.5em minus 0.4em\relax Springer, 2020, pp. 492--511.

\bibitem{liang2022learning}
Y.~Liang, B.~Wang, W.~Ren, J.~Liu, W.~Wang, and W.~Zuo, ``Learning hierarchical dynamics with spatial adjacency for image enhancement,'' in \emph{Proceedings of the 30th ACM International Conference on Multimedia}, 2022, pp. 2767--2776.

\bibitem{lee2013contrast}
C.~Lee, C.~Lee, and C.-S. Kim, ``Contrast enhancement based on layered difference representation of 2d histograms,'' \emph{IEEE transactions on image processing}, vol.~22, no.~12, pp. 5372--5384, 2013.

\bibitem{ma2015perceptual}
K.~Ma, K.~Zeng, and Z.~Wang, ``Perceptual quality assessment for multi-exposure image fusion,'' \emph{IEEE Transactions on Image Processing}, vol.~24, no.~11, pp. 3345--3356, 2015.

\bibitem{huynh2008scope}
Q.~Huynh-Thu and M.~Ghanbari, ``Scope of validity of psnr in image/video quality assessment,'' \emph{Electronics letters}, vol.~44, no.~13, pp. 800--801, 2008.

\bibitem{wang2004image}
Z.~Wang, A.~C. Bovik, H.~R. Sheikh, and E.~P. Simoncelli, ``Image quality assessment: from error visibility to structural similarity,'' \emph{IEEE transactions on image processing}, vol.~13, no.~4, pp. 600--612, 2004.

\bibitem{mittal2012making}
A.~Mittal, R.~Soundararajan, and A.~C. Bovik, ``Making a “completely blind” image quality analyzer,'' \emph{IEEE Signal processing letters}, vol.~20, no.~3, pp. 209--212, 2012.

\bibitem{yang2020advancing}
W.~Yang, Y.~Yuan, W.~Ren, J.~Liu, W.~J. Scheirer, Z.~Wang, T.~Zhang, Q.~Zhong, D.~Xie, S.~Pu \emph{et~al.}, ``Advancing image understanding in poor visibility environments: A collective benchmark study,'' \emph{IEEE Transactions on Image Processing}, vol.~29, pp. 5737--5752, 2020.

\bibitem{loh2019getting}
Y.~P. Loh and C.~S. Chan, ``Getting to know low-light images with the exclusively dark dataset,'' \emph{Computer Vision and Image Understanding}, vol. 178, pp. 30--42, 2019.

\end{thebibliography}

\vfill

\end{document}